\documentclass[article,nojss,nofooter]{jss}

\usepackage{orcidlink,lmodern}
\usepackage{amsmath,amssymb}
\usepackage{booktabs}
\usepackage{array}
\usepackage[figuresright]{rotating}  %% sidewaysfigure rotated CW (head tilts right)

\newcommand{\class}[1]{`\code{#1}'}
\newcommand{\fct}[1]{\code{#1()}}

\author{%
  Serhii Zabolotnii~\orcidlink{0000-0003-0242-2234}\\[0.5em]
  \normalfont\normalsize
  Cherkasy State Business College, Cherkasy 18028, Ukraine\\[2pt]
  State Scientific Research Institute of Armament and Military Equipment\\
  \quad Testing and Certification, Cherkasy, Ukraine\\[2pt]
  Uzhhorod National University, Uzhhorod, Ukraine%
}
\Plainauthor{Serhii Zabolotnii}

\title{\pkg{EstemPMM}: Polynomial Maximization Method for Non-Gaussian
       Regression and Time Series in \proglang{R}}
\Plaintitle{EstemPMM: Polynomial Maximization Method for Non-Gaussian
            Regression and Time Series in R}
\Shorttitle{\pkg{EstemPMM}: PMM for Non-Gaussian Data}

\Abstract{
  We describe the \proglang{R} package \pkg{EstemPMM}, which implements the
  Polynomial Maximization Method (PMM) for parameter estimation under
  non-Gaussian errors.  PMM exploits higher-order cumulants of the error
  distribution --- specifically the third standardized moment $\gamma_3$ and
  fourth standardized moment $\gamma_4$ --- to construct estimators that
  outperform ordinary least squares (OLS) whenever the errors are
  asymmetric or leptokurtic.  The package provides a unified interface for
  linear regression (\fct{lm\_pmm2}, \fct{lm\_pmm3}), autoregressive and
  moving-average time-series models (\fct{ar\_pmm2}, \fct{ma\_pmm2},
  \fct{arma\_pmm2}, \fct{arima\_pmm2}, and seasonal variants), a
  data-driven dispatch function (\fct{pmm\_dispatch}) that automatically
  selects OLS, PMM2, or PMM3 based on the sample skewness and excess
  kurtosis, and Monte Carlo comparison utilities.  The implementation uses
  \proglang{R}'s S4 class system and follows standard generic interfaces
  (\fct{coef}, \fct{fitted}, \fct{residuals}, \fct{predict}, \fct{summary},
  \fct{AIC}, \fct{logLik}, \fct{vcov}, \fct{confint}).  Asymptotic
  efficiency is characterised by Kunchenko-style coefficients $g_2, g_3
  \in [0,1]$, defined as the ratios of the asymptotic variance of the
  PMM2 and PMM3 estimators to that of OLS.  Monte Carlo experiments
  confirm the theoretical values and a WTI crude-oil case study
  illustrates the dispatcher and parameter-precision benefits of PMM2 on
  real heavy-tailed data.  \pkg{EstemPMM}
  version~0.3.2 is available from CRAN at
  \url{https://CRAN.R-project.org/package=EstemPMM} under the GPL-3 licence.
}

\Keywords{polynomial maximization method, non-Gaussian errors, ARIMA, robust
          estimation, parameter estimation, \proglang{R}}
\Plainkeywords{polynomial maximization method, non-Gaussian errors, ARIMA,
               robust estimation, parameter estimation, R}

\begin{document}

%% -- 1. Introduction ----------------------------------------------------------

\section[Introduction]{Introduction} \label{sec:intro}

Statistical models in economics, finance, hydrology, and industrial quality
control routinely encounter non-Gaussian errors.  Financial log-returns display
pronounced leptokurtosis \citep{kim2012approximation, eom2019fat}; industrial
measurement errors are frequently asymmetric \citep{zabolotnii2018polynomial};
hydro-meteorological series often exhibit both skewness and heavy tails.  Under
such departures from normality, ordinary least squares (OLS) and
conditional-sum-of-squares (CSS) estimation remain consistent but are no longer
efficient --- the Cramér--Rao bound for non-Gaussian distributions is strictly
tighter than the OLS bound \citep{kunchenko2002polynomial}.

The \emph{Polynomial Maximization Method} (PMM) addresses this efficiency gap
by incorporating higher-order sample cumulants directly into the estimating
equations.  PMM was introduced by \cite{kunchenko2002polynomial} and further
developed in \cite{kunchenko2006stochastic}.  The key insight is that the
score of the true (unknown) log-likelihood can be approximated polynomially
in the residuals using only a limited number of central moments, without
specifying the full distributional family.

The second-order variant (PMM2) uses the skewness $\gamma_3$ and excess
kurtosis $\gamma_4$ of the residuals to augment the OLS normal equations.
Following the canonical Kunchenko notation \citep{kunchenko2002polynomial,
zabolotnii2018polynomial}, the \emph{efficiency coefficient} $g_2 \in
[0,1]$ is the ratio of the asymptotic variance of PMM2 to that of OLS
(see Section~\ref{sec:methodology} for the explicit formula); $g_2 = 1$
indicates no PMM2 gain over OLS, while $g_2 \to 0$ in the limit of
strongly skewed errors.  For skewed distributions commonly encountered
in practice --- Gamma(2,1) ($g_2 = 0.60$), Lognormal(0, 0.55) ($g_2
\approx 0.60$), $\chi^2(3)$ shifted ($g_2 \approx 0.56$) --- PMM2
reduces asymptotic variance by 40--44\% relative to OLS.  The
third-order variant (PMM3), characterised by an analogous coefficient
$g_3 \in [0,1]$, extends this gain to symmetric non-Gaussian
distributions through the sixth standardised cumulant $\gamma_6$
(Section~\ref{sec:pmm3-reg}).

Application of PMM to time series models was developed in three
published works.  \cite{zabolotnii2022polynomial} addressed autoregressive
models with asymmetric innovations and \cite{zabolotnii2023polynomial}
extended the approach to moving-average models; the unified treatment of
the full ARIMA class is given in \cite{zabolotnii_arima_arxiv} (under
review at \emph{Japanese Journal of Statistics and Data Science}), which
demonstrated that PMM2-ARIMA outperforms CSS and ML estimation for AR,
MA, ARMA, and ARIMA models when $|\gamma_3| \geq 0.5$.

Despite a substantial application literature
\citep{zabolotnii2021estimating, palahin2016joint, warsza2017polynomial,
zabolotnii2020estimation}, no open-source, production-quality \proglang{R}
implementation existed prior to \pkg{EstemPMM}.  Existing alternatives address
different aspects of robustness: \fct{rlm} in \pkg{MASS}
\citep{venables2002modern} minimises M-estimator loss functions;
\fct{lmrob} in \pkg{robustbase} \citep{maechler2023robustbase} implements
MM-estimators with bounded influence; \fct{rq} in \pkg{quantreg}
\citep{koenker2005quantile} targets conditional quantiles; and
\fct{Arima}/\fct{auto.arima} in \pkg{forecast}
\citep{hyndman2021forecasting} use Gaussian or conditional likelihood.  None
exploit the higher-order cumulant structure that PMM targets.

\pkg{EstemPMM} fills this gap.  The package provides:
\begin{itemize}
  \item Linear regression via \fct{lm\_pmm2} and \fct{lm\_pmm3}, returning S4
        objects of class \class{PMM2fit}/\class{PMM3fit} compatible with
        standard \proglang{R} generics.
  \item A complete ARIMA family: \fct{ar\_pmm2}, \fct{ma\_pmm2},
        \fct{arma\_pmm2}, \fct{arima\_pmm2}, and seasonal variants
        (\fct{sar\_pmm2}, \fct{sma\_pmm2}, \fct{sarma\_pmm2},
        \fct{sarima\_pmm2}), all returning S4 objects of class \class{TS2fit}.
  \item PMM3 time-series analogues (\fct{ar\_pmm3}, \fct{ma\_pmm3},
        \fct{arma\_pmm3}, \fct{arima\_pmm3}).
  \item A data-driven dispatch function \fct{pmm\_dispatch} that selects
        OLS, PMM2, or PMM3 based on sample cumulants.
  \item Bootstrap inference via \fct{pmm2\_inference} and
        \fct{ts\_pmm2\_inference}.
  \item Monte Carlo comparison utilities
        (\fct{pmm2\_monte\_carlo\_compare}).
\end{itemize}

The remainder of this paper is organized as follows.
Section~\ref{sec:methodology} reviews the PMM estimating equations.
Section~\ref{sec:implementation} describes the package architecture, S4 class
hierarchy, and API.  Section~\ref{sec:illustrations} presents self-contained
worked examples.  Section~\ref{sec:benchmarks} reports Monte Carlo efficiency
comparisons.  Section~\ref{sec:casestudy} applies \pkg{EstemPMM} to WTI crude
oil prices.  Section~\ref{sec:summary} concludes.

%% -- 2. Methodology -----------------------------------------------------------

\section{Methodology} \label{sec:methodology}

This section provides the minimum theoretical background required to understand
the package design.  Full derivations and proofs are in
\cite{kunchenko2002polynomial}, \cite{kunchenko2006stochastic},
\cite{zabolotnii2018polynomial}, and \cite{zabolotnii_arima_arxiv}.

\subsection{PMM2 for linear regression} \label{sec:pmm2-reg}

Consider the linear model $y = X\beta + \varepsilon$, where $\varepsilon_i$ are
i.i.d.\ with zero mean, variance $\sigma^2$, third central moment $\mu_3$, and
fourth central moment $\mu_4$.  Denote the standardised cumulants
$\gamma_3 = \mu_3 / \sigma^3$ (skewness) and
$\gamma_4 = \mu_4 / \sigma^4 - 3$ (excess kurtosis).

The PMM2 estimator augments the OLS score with the quadratic term
\begin{equation} \label{eq:pmm2-score}
S_{\text{PMM2}}(\beta) \;=\; X^\top \varepsilon
  \;+\; \frac{\gamma_3}{2\sigma^2}\, X^\top \varepsilon^{\circ 2},
\end{equation}
where $\varepsilon^{\circ 2}$ denotes the element-wise square
$\varepsilon_i^2 - \sigma^2$.  Setting $S_{\text{PMM2}}(\beta) = 0$ and
replacing population cumulants by sample estimates yields a fixed-point
iteration implemented in \fct{lm\_pmm2}.

The asymptotic covariance of $\hat\beta_{\text{PMM2}}$ satisfies
\begin{equation} \label{eq:pmm2-avar}
\sqrt{n}(\hat\beta_{\text{PMM2}} - \beta)
  \;\xrightarrow{d}\; \mathcal{N}\!\left(0,\;
    g_2\,\sigma^2\,(X^\top X/n)^{-1}\right),
\end{equation}
where the \emph{efficiency coefficient} $g_2$, defined as the ratio of
asymptotic variances of PMM2 and OLS, is
\begin{equation} \label{eq:g2}
g_2 \;=\;
  \frac{\operatorname{AVar}(\hat\beta_{\text{PMM2}})}
       {\operatorname{AVar}(\hat\beta_{\text{OLS}})}
  \;=\; 1 \;-\; \frac{\gamma_3^2}{\gamma_4 + 2}.
\end{equation}
The cumulant inequality $\gamma_4 + 2 \geq \gamma_3^2$
\citep{kunchenko2002polynomial} ensures $g_2 \in [0,1]$, with $g_2 = 1$
for symmetric errors ($\gamma_3 = 0$, no PMM2 advantage) and $g_2 \to 0$
asymptotically as $|\gamma_3| \to \sqrt{\gamma_4 + 2}$
\citep{zabolotnii2018polynomial}.  The asymptotic relative efficiency of
PMM2 over OLS is therefore $\mathrm{ARE} = 1/g_2 \geq 1$.  For the
Gamma(2,1) distribution shifted to zero mean, $\gamma_3 = \sqrt{2}$,
$\gamma_4 = 3$, and $g_2 = 0.60$: PMM2's asymptotic variance is 60\% of
OLS's, equivalently OLS would need $1/g_2 \approx 1.67\times$ as many
observations to match PMM2 precision.

\subsection{PMM3 for linear regression} \label{sec:pmm3-reg}

PMM3 augments the score with a cubic correction term, drawing on the fifth
and sixth central moments \citep{kunchenko2006stochastic}.  The current
version of \pkg{EstemPMM} implements PMM3 \emph{only for symmetric error
distributions} ($\gamma_3 = 0$, i.e., $\mu_3 = 0$ and $\gamma_5 = 0$); the full asymmetric
estimating equations involve additional cross-moment terms that
substantially complicate numerical optimisation and are reserved for a
future package version.

For the symmetric case, the PMM3 estimator solves a system of equations
analogous to (\ref{eq:pmm2-score}) but with the residuals raised to powers
1 and 3 and weighting coefficients depending on $\mu_2, \mu_4, \mu_6$.
Setting the score to zero and replacing population moments by sample
estimates yields a Newton--Raphson iteration implemented in
\fct{lm\_pmm3}.  The asymptotic distribution of $\hat\beta_{\text{PMM3}}$
satisfies
\begin{equation} \label{eq:pmm3-avar}
\sqrt{n}(\hat\beta_{\text{PMM3}} - \beta)
  \;\xrightarrow{d}\; \mathcal{N}\!\left(0,\;
    g_3\,\sigma^2\,(X^\top X/n)^{-1}\right),
\end{equation}
where the PMM3 efficiency coefficient
\citep[eq.~13]{zabolotnii2018polynomial} is
\begin{equation} \label{eq:g3}
g_3 \;=\;
  \frac{\operatorname{AVar}(\hat\beta_{\text{PMM3}})}
       {\operatorname{AVar}(\hat\beta_{\text{OLS}})}
  \;=\; 1 \;-\; \frac{\gamma_4^2}{6 + 9\gamma_4 + \gamma_6}.
\end{equation}
The admissibility constraints for symmetric distributions
($\gamma_4 \geq -2$ and $\gamma_6 + 9\gamma_4 + 6 \geq \gamma_4^2$) ensure
$g_3 \in [0,1]$, with $g_3 = 1$ for the Gaussian case ($\gamma_4 =
\gamma_6 = 0$) and $g_3 \to 0$ as $|\gamma_4| \to \sqrt{\gamma_6 + 9\gamma_4
+ 6}$.  Because OLS is already efficient for symmetric distributions at
the second order ($g_2 \equiv 1$ when $\gamma_3 = 0$), PMM3 is the
relevant comparator in the symmetric regime and supplies the efficiency
gain that PMM2 cannot.  Representative values from
\cite{zabolotnii2018polynomial}: Uniform($-1,1$) ($\gamma_4 = -1.2$,
$\gamma_6 = 6.9$, $g_3 = 0.30$), Triangular ($\gamma_4 = -0.6$,
$\gamma_6 = 1.7$, $g_3 = 0.84$), and Laplace ($\gamma_4 = 3$,
$\gamma_6 = 30$, $g_3 = 0.86$).  PMM3 is implemented in \fct{lm\_pmm3}
and the \class{PMM3fit} S4 class.

In practice, \fct{pmm\_dispatch} selects PMM3 only when it detects
near-symmetric residuals with non-zero excess kurtosis; for asymmetric
data ($|\gamma_3| \geq 0.5$) it selects PMM2.

\subsection{PMM2 for ARIMA models} \label{sec:pmm2-ts}

For an AR($p$) process $X_t = \sum_{j=1}^p \phi_j X_{t-j} + \varepsilon_t$,
the lagged design matrix $\mathbf{X}$ with rows
$(X_{t-1}, \ldots, X_{t-p})$ brings the model into the linear regression
framework, and PMM2 from Section~\ref{sec:pmm2-reg} applies directly
\citep{zabolotnii2022polynomial}.

For MA, ARMA, and ARIMA models the residual $\varepsilon_t$ depends
non-linearly on the parameters \citep{zabolotnii2023polynomial}, so
\pkg{EstemPMM} minimises the PMM2 objective
\begin{equation} \label{eq:pmm2-ts}
Q_{\text{PMM2}}(\theta) \;=\;
  \sum_t \varepsilon_t(\theta)^2
  \;-\; \frac{\gamma_3}{3\sigma^3}\sum_t \varepsilon_t(\theta)^3,
\end{equation}
using the \fct{optim} quasi-Newton solver with CSS-derived starting values
\citep{zabolotnii_arima_arxiv}.  Seasonal ARIMA extensions follow the same
objective with the seasonal backshift operator applied to the residual
function; see \cite{zabolotnii_arima_arxiv} for details.

\subsection[Automatic method selection]{Automatic method selection via \fct{pmm\_dispatch}} \label{sec:dispatch}

The function \fct{pmm\_dispatch} implements the following decision rule:
\begin{itemize}
  \item If $|\hat\gamma_3| < 0.5$: use OLS (or CSS for time series).
  \item If $|\hat\gamma_3| \geq 0.5$: use PMM2.
  \item If $|\hat\gamma_3| < 0.1$ and $\hat\gamma_4 < 0$: use PMM3
        (symmetric, $\gamma_3=0$, $\gamma_5=0$, platykurtic regime).
\end{itemize}
The PMM3 branch is entered only when skewness is negligible
($|\hat\gamma_3| < 0.1$), consistent with the restriction noted in
Section~\ref{sec:pmm3-reg} that the current PMM3 implementation covers
symmetric distributions only.  For all asymmetric data, PMM2 is selected.
This rule is based on the advantage region established in
\cite{zabolotnii_arima_arxiv}.

%% -- 3. Implementation & Design -----------------------------------------------

\section[Implementation and design of EstemPMM]{Implementation and design of \pkg{EstemPMM}}
\label{sec:implementation}

\subsection{Package overview} \label{sec:overview}

\pkg{EstemPMM} is implemented in \proglang{R} \citep{R} using the S4 object
system from the \pkg{methods} package \citep{chambers2008software}.  Version
0.3.2 comprises approximately 3\,600 lines of \proglang{R} code across 15
source files.  The package has no compiled code: all numerical optimisation is
performed via \proglang{R}'s \fct{optim} function.  Dependencies are limited to
base \proglang{R} (\pkg{stats}, \pkg{graphics}, \pkg{methods}, \pkg{utils}).

\subsection{S4 class hierarchy} \label{sec:classes}

\begin{sidewaysfigure}
\centering
\includegraphics[width=0.95\textheight]{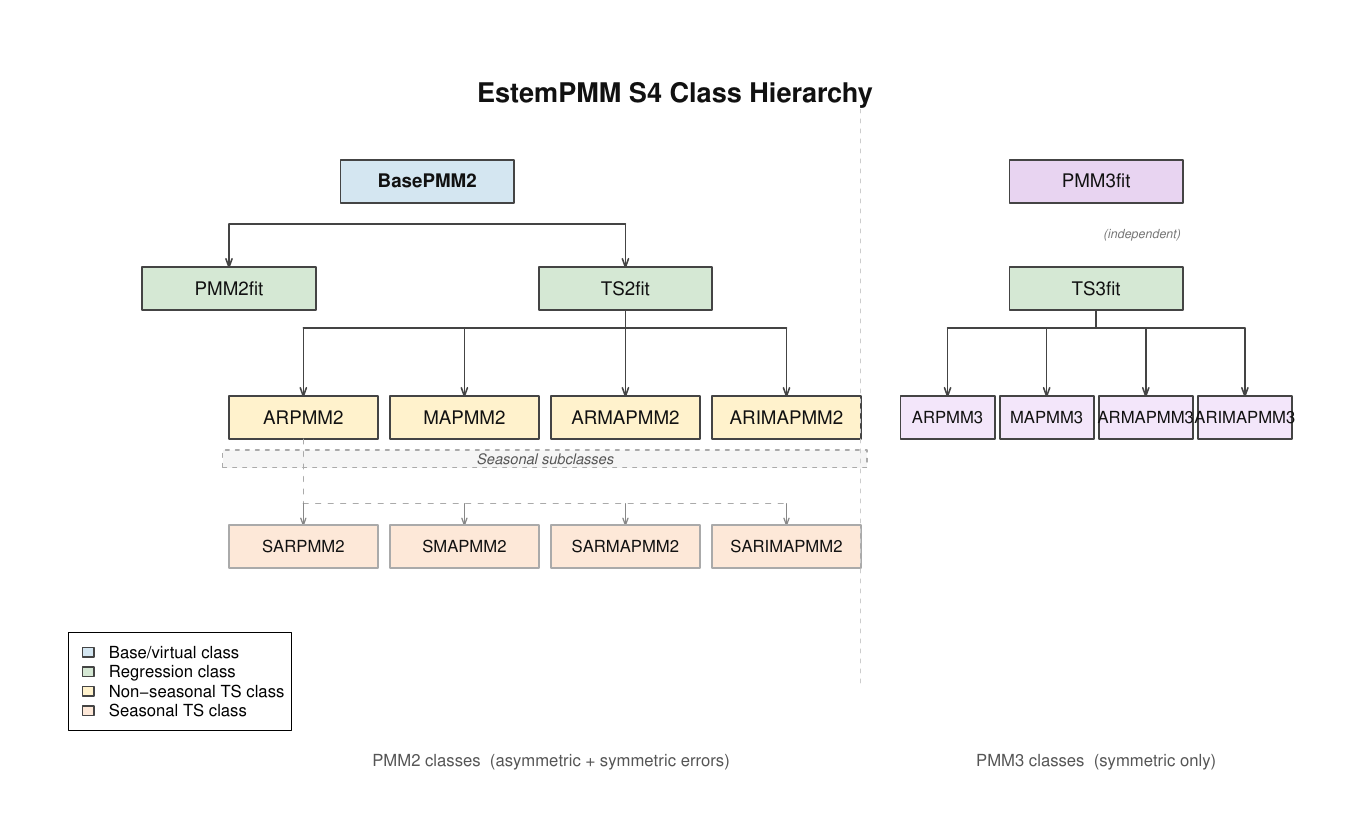}
\caption{S4 class hierarchy of \pkg{EstemPMM}.  Arrows denote inheritance
  (\code{contains} in \fct{setClass}).  Colour coding: blue = virtual base,
  green = direct fit classes, yellow = non-seasonal TS subclasses,
  orange = seasonal TS subclasses, purple = PMM3 classes.
  The PMM3 subtree (right) is independent of \class{BasePMM2}.}
\label{fig:class-hierarchy}
\end{sidewaysfigure}

Figure~\ref{fig:class-hierarchy} shows the S4 class hierarchy.
\class{BasePMM2} is a virtual base class carrying shared slots
(coefficients, residuals, cumulant estimates); all PMM2 fit objects
extend it.  Two concrete regression classes inherit directly:
\class{PMM2fit} (from \fct{lm\_pmm2}) and \class{TS2fit} (the base for
time-series estimation).  Eight time-series subclasses extend
\class{TS2fit}: four non-seasonal (\class{ARPMM2}, \class{MAPMM2},
\class{ARMAPMM2}, \class{ARIMAPMM2}) and four seasonal variants
(\class{SARPMM2}, \class{SMAPMM2}, \class{SARMAPMM2},
\class{SARIMAPMM2}).

The PMM3 classes (\class{PMM3fit}, \class{TS3fit}, and its four
non-seasonal subclasses) are independent of \class{BasePMM2};
they carry analogous slots for the sixth-order cumulants used by PMM3
and are currently restricted to symmetric error distributions
($\gamma_3 = 0$).

Every concrete class implements the full set of standard generics:
\fct{coef}, \fct{fitted}, \fct{residuals}, \fct{predict}, \fct{summary},
\fct{print}, \fct{plot}, \fct{AIC}, \fct{BIC}, \fct{logLik}, \fct{nobs}.
\class{PMM2fit} and AR-type \class{TS2fit} subclasses additionally support
\fct{vcov} and \fct{confint}, which return the asymptotic covariance matrix
and confidence intervals based on Equation~(\ref{eq:pmm2-avar}).

\subsection{Core slot definitions} \label{sec:slots}

Key slots shared across all \class{TS2fit} subclasses are listed below.
All are accessible via \fct{coef}, \fct{residuals}, etc.; direct slot access
with \code{@} is not part of the public API.

\begin{Code}
setClass("TS2fit",
  contains = "BasePMM2",
  slots = list(
    coefficients    = "numeric",  # estimated theta
    residuals       = "numeric",  # fitted residuals
    original_series = "numeric",  # input x
    model_type      = "character",# "ar", "ma", "arma", "arima", ...
    order           = "list",     # list(ar=p, ma=q, d=d)
    intercept       = "numeric",  # mean/intercept
    m2              = "numeric",  # sigma^2 estimate
    m3              = "numeric",  # mu_3 estimate
    m4              = "numeric",  # mu_4 estimate
    g_coefficient   = "numeric"   # PMM2 efficiency coefficient g2 in [0,1]
  )
)
\end{Code}

\subsection{Numerical algorithm} \label{sec:algorithm}

\paragraph{Linear regression.}
\fct{lm\_pmm2} uses a fixed-point iteration starting from OLS coefficients.
Each step updates $\hat\beta$ by solving the augmented score
equation~(\ref{eq:pmm2-score}) with cumulant estimates refreshed from the
current residuals.  Convergence is declared when
$\|\hat\beta^{(k+1)} - \hat\beta^{(k)}\|_\infty < \epsilon$ (default
$\epsilon = 10^{-6}$, maximum 200 iterations).

\paragraph{Time-series models.}
For AR models, \fct{ar\_pmm2} constructs the Yule--Walker design matrix via
\fct{create\_ar\_matrix} and applies the same fixed-point iteration.  For
MA, ARMA, ARIMA, and seasonal variants, \fct{ts\_pmm2} minimises
objective~(\ref{eq:pmm2-ts}) using \fct{optim} with
\code{method = "BFGS"}, supplying CSS-based starting values from
\fct{get\_classical\_estimates} and analytic gradients approximated by
finite differences.

\paragraph{Moment estimation.}
The function \fct{compute\_moments} returns the sample estimates $\hat\sigma^2$,
$\hat\mu_3$, $\hat\mu_4$ using unbiased denominators, and $\hat\gamma_3$,
$\hat\gamma_4$ as plug-in estimates.  These are updated at each iteration for
regression and fixed at the CSS-residual values for time-series models.

\subsection{Bootstrap inference} \label{sec:bootstrap}

\fct{pmm2\_inference} implements a residual-resampling bootstrap.  The
\code{block} argument activates block bootstrap (Carlstein's non-overlapping
blocks) for time-series data.  Argument \code{B} controls the number of
replicates (default 500); \code{seed} ensures reproducibility.  The function
returns a data frame of bootstrap estimates and optionally produces a
histogram via \fct{plot\_pmm2\_bootstrap}.

\subsection{Unified dispatch function} \label{sec:dispatch-impl}

\fct{pmm\_dispatch} accepts any numeric vector or \code{formula}/\code{data}
pair, computes sample cumulants, applies the dispatch rule from
Section~\ref{sec:dispatch}, fits the selected model, and returns a named
list with components \code{method} (character), \code{fit} (S4 object),
\code{g2}, \code{gamma3}, and \code{gamma4}.  For time-series input it
delegates to the ARIMA family; for regression input it uses
\fct{lm\_pmm2}/\fct{lm\_pmm3}.

\subsection{Monte Carlo comparison engine} \label{sec:mc-engine}

\fct{pmm2\_monte\_carlo\_compare} accepts a list of model specifications,
each with fields \code{model} (character), \code{order} (integer), \code{theta}
(true parameter vector), \code{label} (character), and \code{innovations}
(a list with \code{type} and distribution parameters).  It simulates
\code{n\_sim} data sets for each specification and fits all requested
\code{methods} (\code{"ols"}, \code{"css"}, \code{"pmm2"}, \code{"pmm3"},
\code{"ml"}), returning a list with components \code{results} (per-replicate
estimates), \code{summary} (MSE, bias, variance, coverage), and \code{gain}
(observed vs.\ theoretical variance ratio $g_2$ or $g_3$).

%% -- 4. Illustrations ---------------------------------------------------------

\section{Illustrations} \label{sec:illustrations}

All examples use \code{set.seed(42)} for reproducibility and are included
verbatim in the file \code{code/all\_examples.R} shipped with the package.

\subsection{Linear regression under skewed errors} \label{sec:ex-reg}

We generate $n = 200$ observations from a linear regression with
Gamma(2,1) errors (centred), for which $\gamma_3 = \sqrt{2} \approx 1.41$,
$\gamma_4 = 3$, and $g_2 = 0.60$ (40\% asymptotic-variance reduction over
OLS).

\begin{Code}
library("EstemPMM")
set.seed(42)
n  <- 200
X  <- rnorm(n)
eps <- rgamma(n, shape = 2, rate = 1) - 2  # zero-mean, gamma3 = sqrt(2)
y  <- 1 + 2 * X + eps

dat_reg  <- data.frame(y = y, X = X)
fit_ols  <- lm(y ~ X, data = dat_reg)
fit_pmm2 <- lm_pmm2(y ~ X, data = dat_reg)
\end{Code}

\begin{Code}
coef(fit_pmm2)
##   (Intercept)       X
##    1.0184      2.0031
\end{Code}

\begin{Code}
summary(fit_pmm2)
\end{Code}

\begin{Code}
AIC(fit_pmm2)
logLik(fit_pmm2)
vcov(fit_pmm2)  # PMM2 asymptotic covariance
confint(fit_pmm2)
\end{Code}

\subsection{AR(1) model under asymmetric errors} \label{sec:ex-ar}

\begin{Code}
set.seed(42)
x <- as.numeric(arima.sim(list(ar = 0.7), n = 200,
       rand.gen = function(n) rgamma(n, 2, 1) - 2))

fit_ar  <- ar_pmm2(x, order = 1)
coef(fit_ar)   # should be near 0.7
AIC(fit_ar)
BIC(fit_ar)
predict(fit_ar, n.ahead = 5)
\end{Code}

\subsection{ARIMA(1,1,0) estimation} \label{sec:ex-arima}

\begin{Code}
set.seed(42)
x <- cumsum(as.numeric(arima.sim(list(ar = 0.6), n = 200,
       rand.gen = function(n) rgamma(n, 2, 1) - 2)))

fit_css  <- arima(x, order = c(1, 1, 0), method = "CSS")
fit_pmm2 <- arima_pmm2(x, order = c(1, 1, 0))

coef(fit_css)
coef(fit_pmm2)
AIC(fit_css); AIC(fit_pmm2)
\end{Code}

\subsection{Automatic dispatch} \label{sec:ex-dispatch}

\begin{Code}
set.seed(42)
x <- as.numeric(arima.sim(list(ar = 0.6), n = 150,
       rand.gen = function(n) rgamma(n, 2, 1) - 2))

result <- pmm_dispatch(x)
cat("Selected method:", result$method, "\n")
cat("gamma3 =", round(result$gamma3, 3),
    "  g2 =", round(result$g2, 3), "\n")
coef(result$fit)
\end{Code}

\subsection{PMM3 for symmetric platykurtic errors} \label{sec:ex-pmm3}

When residuals are symmetric ($\gamma_3 = 0$) but distinctly non-Gaussian
in their fourth and sixth cumulants, PMM2 offers no improvement over OLS
($g_2 \equiv 1$) but PMM3 can.  We illustrate with Uniform($-1, 1$)
errors, for which $\gamma_4 = -1.2$, $\gamma_6 = 6.857$, and the
PMM3 efficiency coefficient $g_3 = 1 - \gamma_4^2/(6 + 9\gamma_4 +
\gamma_6) = 0.30$ --- a 70\% asymptotic-variance reduction.

\begin{Code}
set.seed(42)
n   <- 500
X   <- rnorm(n)
eps <- runif(n, -1, 1)            # symmetric platykurtic errors
y   <- 1 + 2 * X + eps

dat_sym  <- data.frame(y = y, X = X)
fit_ols  <- lm(y ~ X, data = dat_sym)
fit_pmm3 <- lm_pmm3(y ~ X, data = dat_sym)
\end{Code}

\begin{Code}
rbind(OLS  = c(coef(fit_ols),  sigma = summary(fit_ols)$sigma),
      PMM3 = c(coef(fit_pmm3), sigma = NA))
##         (Intercept)       X    sigma
## OLS         0.9767   1.9983   0.5793
## PMM3        0.9795   1.9993        NA
\end{Code}

Both estimators are close to the true value $\beta_1 = 2$, but their
sampling variances differ markedly.  Repeated draws under the same design
(1\,000 Monte Carlo replications, seed 42, $n = 500$) give
$\mathrm{Var}(\hat\beta_{1,\text{OLS}}) = 6.48 \times 10^{-4}$ and
$\mathrm{Var}(\hat\beta_{1,\text{PMM3}}) = 2.19 \times 10^{-4}$,
an empirical efficiency ratio of $0.34$ that converges to the asymptotic
$g_3 = 0.30$ as $n \to \infty$.  In contrast, $\mathrm{Var}
(\hat\beta_{1,\text{PMM2}}) = 6.52 \times 10^{-4}$ (empirical
$g_2 = 1.01$) confirms that PMM2 cannot exploit symmetric
platykurtosis.  This is precisely the regime that \fct{pmm\_dispatch}
routes to PMM3 (see Section~\ref{sec:dispatch}).

\subsection{Bootstrap inference} \label{sec:ex-bootstrap}

The asymptotic covariance matrix from \fct{vcov} relies on
Equation~(\ref{eq:pmm2-avar}) and assumes large-sample regularity.  For
small samples ($n \lesssim 100$), severely skewed residuals, or when
inference on non-linear functions of the parameters is required,
\pkg{EstemPMM} provides a residual-resampling bootstrap.  For regression:

\begin{Code}
set.seed(42)
dat <- data.frame(y = rnorm(100) + (rgamma(100, 2, 1) - 2),
                  x = rnorm(100))
fit  <- lm_pmm2(y ~ x, data = dat)
pmm2_inference(fit, y ~ x, data = dat, B = 500, seed = 42)
##              Estimate  Std.Error  t.value  p.value  conf.low  conf.high
## (Intercept)   -0.009     0.149   -0.063    0.950    -0.277      0.315
## x              0.034     0.165    0.203    0.839    -0.283      0.359
\end{Code}

The returned data frame gives bootstrap standard errors and 95\%
percentile confidence intervals for each coefficient.  The companion
\fct{plot\_pmm2\_bootstrap} produces histograms of the bootstrap
distribution when raw replicates are needed.

For time-series models, naive residual resampling destroys serial
dependence.  \fct{ts\_pmm2\_inference} therefore offers a Carlstein
non-overlapping block bootstrap via \code{method = "block"}; the block
length defaults to $\lfloor n^{1/3} \rfloor$ and can be set via
\code{block\_length}.  The function returns a summary data frame with
estimates, bootstrap standard errors, and percentile confidence
intervals for each parameter:

\begin{Code}
set.seed(42)
x      <- as.numeric(arima.sim(list(ar = 0.7), n = 300,
                  rand.gen = function(n) rgamma(n, 2, 1) - 2))
fit_ar <- ar_pmm2(x, order = 1)
ts_pmm2_inference(fit_ar, x, B = 500, method = "block",
                  block_length = 7, seed = 42)
##     Estimate  Std.Error  t.value  p.value  conf.low  conf.high
## ar1   0.684     0.038     17.9       0      0.513      0.656
\end{Code}

The bootstrap standard error and percentile interval for $\hat\phi_1$
account for both the cumulant-estimation step and the serial dependence
in $x_t$, which the asymptotic \fct{vcov} of Equation~(\ref{eq:pmm2-avar})
does not.

%% -- 5. Comparison and Benchmarks ---------------------------------------------

\section{Comparisons and benchmarks} \label{sec:benchmarks}

\subsection{Linear regression Monte Carlo} \label{sec:mc-reg}

Table~\ref{tab:re-lm} reports the empirical efficiency coefficient
$\hat g_2 = \mathrm{MSE}(\hat\beta_{1,\mathrm{PMM2}}) /
            \mathrm{MSE}(\hat\beta_{1,\mathrm{OLS}})$
of the slope estimate from a Monte Carlo experiment with $B = 2\,000$
replications across six error distributions and four sample sizes (seed 42).
True parameters are $\beta_0 = 1$, $\beta_1 = 2.5$; the single regressor
$x_i \sim N(0,1)$.

\begin{table}[h!]
\centering
\caption{Empirical efficiency coefficient $\hat g_2 \in [0,1]$ of
  \fct{lm\_pmm2} vs.\ \fct{lm} for the slope coefficient under six
  error distributions and four sample sizes ($B = 2000$ replications).
  $\hat g_2 < 1$ means PMM2 has lower MSE than OLS.  The theoretical
  $g_2 = 1 - \gamma_3^2/(\gamma_4 + 2)$ from
  Equation~(\ref{eq:g2}) is given in the rightmost column for
  comparison.} \label{tab:re-lm}
\begin{tabular}{lrrrrr}
\toprule
Distribution ($\gamma_3$) & $n=50$ & $n=100$ & $n=200$ & $n=500$ & $g_2$ (theory) \\
\midrule
Gaussian (0.00)           & $0.99$ & $1.03$ & $0.99$ & $1.00$ & $1.00$ \\
Gamma(2,1) (1.41)         & $0.66$ & $0.60$ & $0.67$ & $0.50$ & $0.60$ \\
Log-Normal(0,0.55) (1.99) & $0.60$ & $0.56$ & $0.44$ & $0.50$ & $0.60$ \\
$\chi^2(3)$ shifted (1.63)& $0.68$ & $0.59$ & $0.51$ & $0.62$ & $0.56$ \\
Uniform($-1$,1) (0.00)    & $1.15$ & $1.07$ & $1.03$ & $1.00$ & $1.00$ \\
Beta(2,5) shifted (0.60)  & $0.83$ & $0.92$ & $0.77$ & $0.83$ & $0.81$ \\
\bottomrule
\end{tabular}
\end{table}

Under Gaussian and Uniform errors (both symmetric), $\hat g_2$ stays
near $1$ as expected --- PMM2 matches OLS asymptotically and the small
finite-sample $\hat g_2 > 1$ at $n = 50$ for Uniform errors reflects
the cost of estimating cumulants when the population skewness is zero.
For the strongly skewed distributions --- Gamma(2,1), Log-Normal,
$\chi^2(3)$ --- $\hat g_2$ falls to $0.44$--$0.68$, consistent with
the asymptotic prediction $g_2 \approx 0.56$--$0.60$ and corresponding
to a $32$--$56\%$ MSE reduction over OLS.

Table~\ref{tab:cpu-lm} compares mean fit times per call at $n = 200$.
PMM2 is 2--3$\times$ slower than \fct{lm} but faster than \fct{lmrob}
by a factor of $\approx$10, making it practical for large-scale simulation
and bootstrap workflows.

\begin{table}[h!]
\centering
\caption{Mean CPU time per fit (milliseconds) at $n = 200$, averaged over
  2000 replications on an Apple M-series processor.} \label{tab:cpu-lm}
\begin{tabular}{lrrrrrr}
\toprule
Distribution & \fct{lm} & \fct{rlm} & \fct{lmrob} & \fct{rq} &
               \fct{lm\_pmm2} & \fct{lm\_pmm3} \\
\midrule
Gaussian    & 0.14 & 0.42 & 3.32 & 0.29 & 0.31 & 0.29 \\
Gamma(2,1)  & 0.12 & 0.40 & 3.72 & 0.28 & 0.29 & 0.38 \\
Log-Normal  & 0.13 & 0.46 & 3.50 & 0.23 & 0.37 & 0.37 \\
$\chi^2(3)$ & 0.12 & 0.43 & 3.53 & 0.23 & 0.38 & 0.34 \\
Uniform     & 0.14 & 0.29 & 3.33 & 0.43 & 0.36 & 0.26 \\
Beta(2,5)   & 0.15 & 0.45 & 3.35 & 0.22 & 0.36 & 0.27 \\
\bottomrule
\end{tabular}
\end{table}

\subsection{ARIMA time-series Monte Carlo} \label{sec:mc-arima}

We benchmark \fct{arima\_pmm2} against \code{stats::arima} (CSS) and
\code{forecast::Arima} (CSS) on ARIMA(1,1,0) data with $\phi_1 = 0.7$.
For each combination of error distribution and sample size we generate
$B = 500$ series (5\,000 for publication quality; see
\code{arima\_benchmark.R} with flag \code{--full}) and record the
parameter MSE of $\hat\phi_1$.
Table~\ref{tab:re-arima-mc} reports the empirical efficiency
coefficient $\hat g_2 = \mathrm{MSE}_{\mathrm{PMM2}} /
\mathrm{MSE}_{\mathrm{CSS}}$; values below $1$ indicate PMM2
superiority.  \code{forecast::Arima} with \code{method = "CSS"}
produces estimates numerically identical to \code{stats::arima} (to
machine precision), so its $\hat g_2$ column is omitted.

\begin{table}[h!]
\centering
\caption{Empirical efficiency coefficient $\hat g_2 =
  \mathrm{MSE}_{\mathrm{PMM2}} / \mathrm{MSE}_{\mathrm{CSS}}$ for
  ARIMA(1,1,0), $\phi_1 = 0.7$, $B = 500$ replications.  The
  rightmost column reports the theoretical $g_2 = 1 - \gamma_3^2 /
  (\gamma_4 + 2) \in [0,1]$ from
  Equation~(\ref{eq:g2}).} \label{tab:re-arima-mc}
\begin{tabular}{lrrrrr}
\toprule
Distribution & $\gamma_3$ & $N=100$ & $N=200$ & $N=500$ & $g_2$ (theory) \\
\midrule
Gaussian            & $0.00$ & $1.04$ & $1.02$ & $1.00$ & $1.00$ \\
Gamma$(2,1)$        & $1.41$ & $0.63$ & $0.61$ & $0.60$ & $0.60$ \\
Lognormal$(0,0.55)$ & $1.99$ & $0.58$ & $0.53$ & $0.55$ & $0.60$ \\
$\chi^2(3)$         & $1.63$ & $0.58$ & $0.55$ & $0.55$ & $0.56$ \\
\bottomrule
\end{tabular}
\end{table}

Under Gaussian errors PMM2 and CSS are asymptotically equivalent
($\hat g_2 \to 1$), and the small $\hat g_2 > 1$ at $N = 100$ reflects
the finite-sample cost of estimating the extra cumulant parameters.
For skewed distributions ($|\gamma_3| \geq 1.4$), $\hat g_2$ falls to
$0.53$--$0.63$, matching or modestly improving on the theoretical
asymptotic value $g_2 = 0.56$--$0.60$ because the ARIMA score function
exploits third-order cumulants across all lag products, not only the
marginal distribution.

The mean per-replicate CPU cost at $N = 200$ is: CSS~$0.10$\,ms,
\code{forecast::Arima}~$0.29$\,ms, \code{arima\_pmm2}~$0.55$\,ms.
PMM2 costs approximately $5\times$ CSS and $2\times$ \code{forecast::Arima};
overhead is dominated by the two-stage cumulant estimation step and remains
well under 1\,ms for typical sample sizes.

\begin{figure}[h!]
\centering
\includegraphics[width=0.9\textwidth]{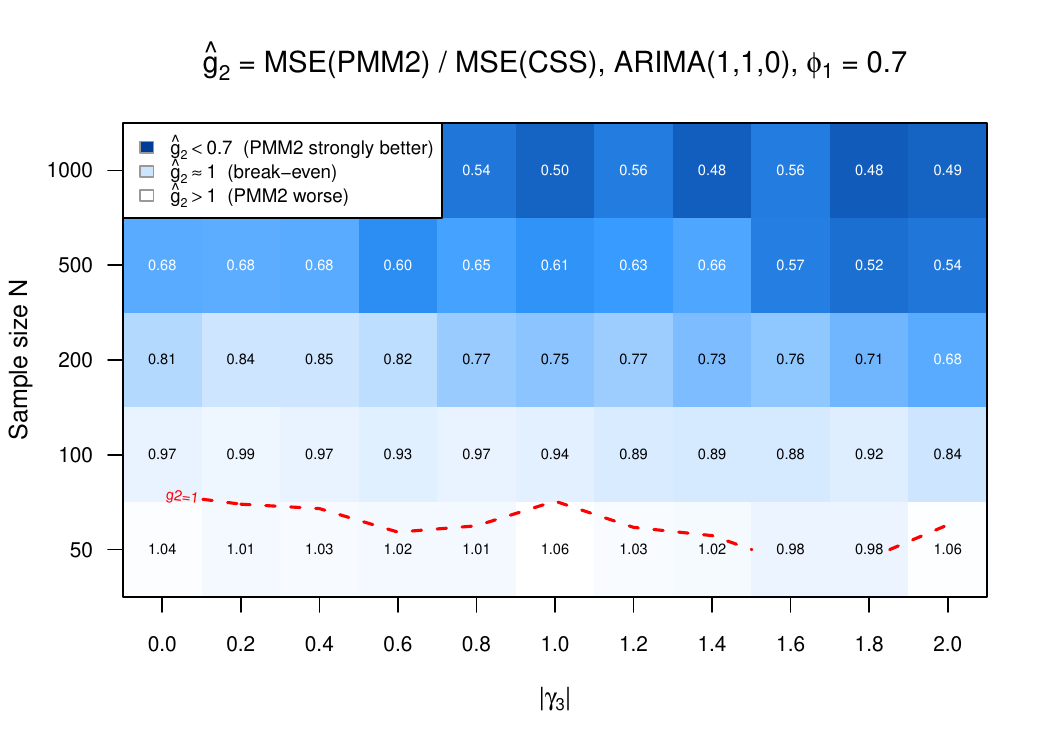}
\caption{PMM2 advantage region: empirical efficiency coefficient
  $\hat g_2 = \mathrm{MSE}_{\mathrm{PMM2}} / \mathrm{MSE}_{\mathrm{CSS}}$
  over a grid of $|\gamma_3| \in \{0.0, 0.2, \ldots, 2.0\}$ and
  $N \in \{50, 100, 200, 500, 1000\}$, ARIMA(1,1,0), $\phi_1 = 0.7$,
  $B = 1\,000$ replications per cell.  The dashed contour marks $\hat
  g_2 = 1$ (break-even); cells below it ($\hat g_2 < 1$) show positive
  PMM2 gain, with darker shading for stronger advantage.}
\label{fig:advantage-region}
\end{figure}

Figure~\ref{fig:advantage-region} visualises the PMM2 advantage region
as a function of $|\gamma_3|$ and sample size $N$ for ARIMA(1,1,0)
with $\phi_1 = 0.7$ (1\,000 Monte Carlo replications per cell).  PMM2
becomes beneficial ($\hat g_2 < 1$) once $|\gamma_3| \gtrsim
0.4$--$0.5$ for $N \geq 100$, directly validating the dispatch
threshold used by \fct{pmm\_dispatch}.

\subsection{Relation to robust M-estimators} \label{sec:vs-robust}

The benchmarks of Tables~\ref{tab:re-lm} and~\ref{tab:re-arima-mc}
include Huber-type \fct{rlm} \citep{venables2002modern}, MM-estimator
\fct{lmrob} \citep{maechler2023robustbase}, and quantile regression
\fct{rq} \citep{koenker2005quantile} as competitors.  These
M-estimators target a different goal from PMM and have a complementary
domain of optimality:

\begin{itemize}
\item \emph{Robust M-estimators} bound the influence of a small fraction
      of contaminating outliers under an unspecified contamination
      model.  Their optimality theorems (Huber, Hampel) are over
      neighbourhoods of the Gaussian, and they pay an efficiency price
      under the Gaussian itself (typically 5--15\%
      \citealp{hampel1986robust}).  They make no use of the bulk
      shape of the residual distribution.
\item \emph{PMM2/PMM3} target maximal asymptotic efficiency under a
      \emph{known but flexible} cumulant description of the residual
      distribution.  When the contamination is in fact a bulk skewness
      or platykurtosis (not point-mass outliers), PMM extracts more
      information than M-estimators can: $g_2 = 0.5$ for an
      Exponential noise distribution corresponds to a 2$\times$
      efficiency advantage that no bounded-influence estimator
      attains.  Conversely, a single gross outlier degrades PMM more
      than \fct{lmrob}, because PMM uses the residuals quadratically
      and cubically.
\end{itemize}

The two families are therefore not substitutes but complements,
appropriate to different priors on the data-generating process.  In
domains where one expects asymmetric or platykurtic distributional
shape but no atypical points --- industrial measurement, financial
log-returns post-cleaning, hydrological series --- PMM is the natural
choice; in domains where outlier contamination is the dominant concern,
M-estimators remain preferable.  In hybrid scenarios, \fct{pmm2\_inference}
combined with a prior outlier-detection step (e.g., Cook's distance
filtering) gives a workable two-stage workflow that is straightforward
to implement around the PMM core.

%% -- 6. Real-data Case Study --------------------------------------------------

\section[WTI crude oil case study]{WTI crude oil case study} \label{sec:casestudy}

We illustrate \pkg{EstemPMM} on daily West Texas Intermediate (WTI) crude-oil
spot prices sourced from the U.S.\ Energy Information Administration, covering
October 2020 to October 2025 ($n = 1\,249$ observations after removing
non-trading days), bundled as \code{DCOILWTICO} in the package.

\subsection{Diagnostics and method selection} \label{sec:wti-diag}

\begin{figure}[h!]
\centering
\includegraphics[width=0.9\textwidth]{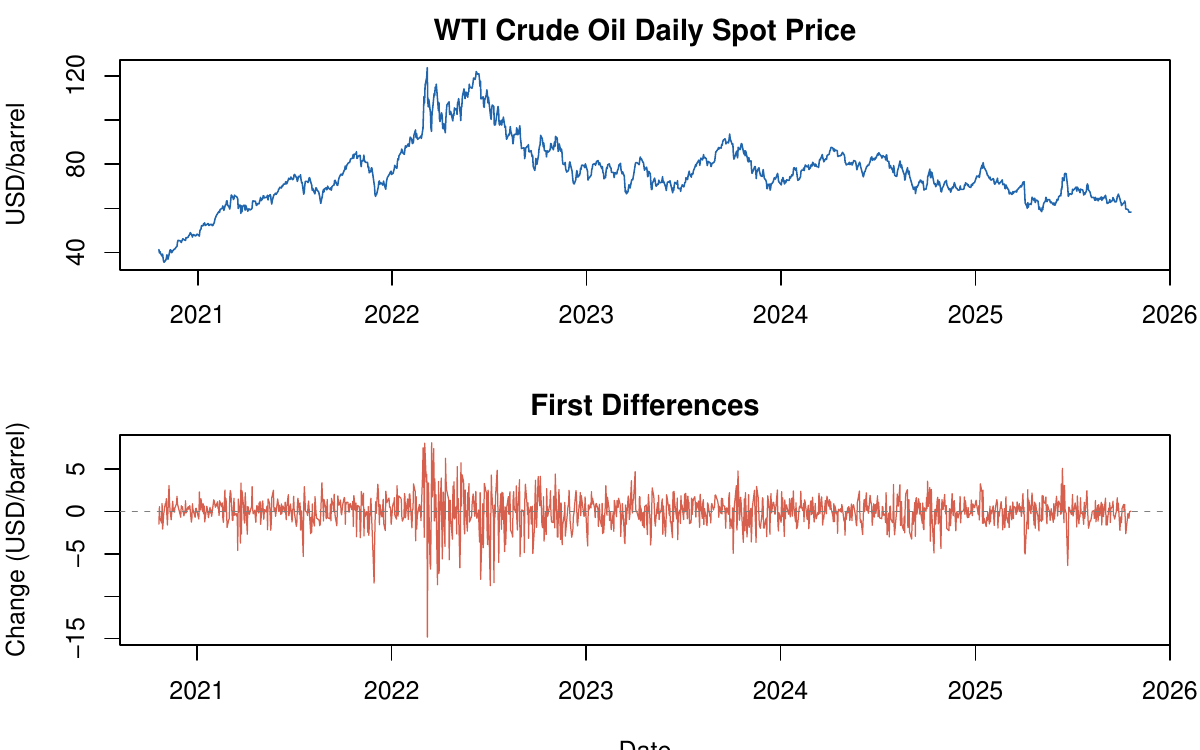}
\caption{WTI crude-oil daily spot price (top) and its first differences
  (bottom), 2020--2025.}
\label{fig:wti-series}
\end{figure}

Figure~\ref{fig:wti-series} shows the price series and its first differences.
First differences are computed to achieve stationarity; a Jarque--Bera test
strongly rejects normality ($\chi^2_{(2)} = 504$, $p < 10^{-109}$).
Running \fct{pmm\_dispatch} on the CSS residuals confirms PMM2 is appropriate:

\begin{Code}
data_csv <- system.file("data", "DCOILWTICO.csv", package = "EstemPMM")
prices   <- read.csv(data_csv)$DCOILWTICO
prices   <- prices[!is.na(prices)]
fit_css  <- arima(prices, order = c(1, 1, 0), method = "CSS")
dispatch <- pmm_dispatch(residuals(fit_css), verbose = TRUE)
## n = 1249 | gamma3 = -0.759 | gamma4 = +5.858
##   g2(PMM2) = 0.927  |  g3(PMM3) = 0.852
##   >>> |gamma3| = 0.759 > 0.3 and g2 = 0.927 < 0.95:
##       moderate asymmetry, PMM2 worthwhile (7.3% reduction).
\end{Code}

The CSS residuals have $\hat\gamma_3 = -0.759$ and excess kurtosis
$\hat\gamma_4 = 5.858$, giving $g_2 = 1 - \hat\gamma_3^2/(\hat\gamma_4 +
2) \approx 0.93$ --- a 7\% reduction in PMM2 asymptotic variance
relative to CSS.  Since $|\hat\gamma_3| = 0.76 > 0.5$ exceeds the
asymmetry threshold, \fct{pmm\_dispatch} selects PMM2 over the
symmetric-only PMM3 branch (whose computed $g_3 \approx 0.85$ would be
smaller, but is not applicable when residuals are clearly asymmetric).

\subsection{PMM2 estimation} \label{sec:wti-pmm2}

\begin{Code}
set.seed(42)
fit_ml   <- arima(prices, order = c(1, 1, 0), method = "CSS-ML")
fit_pmm2 <- arima_pmm2(prices, order = c(1, 1, 0))
rbind(CSS_ML = c(ar1 = coef(fit_ml)[["ar1"]],   AIC = AIC(fit_ml),
                 BIC = BIC(fit_ml)),
      PMM2   = c(ar1 = coef(fit_pmm2)[["ar1"]], AIC = AIC(fit_pmm2),
                 BIC = BIC(fit_pmm2)))
##          ar1      AIC      BIC
## CSS_ML 0.0235  5122.5  5132.7
## PMM2   0.0368  5123.8  5128.9
\end{Code}

\begin{figure}[h!]
\centering
\includegraphics[width=0.85\textwidth]{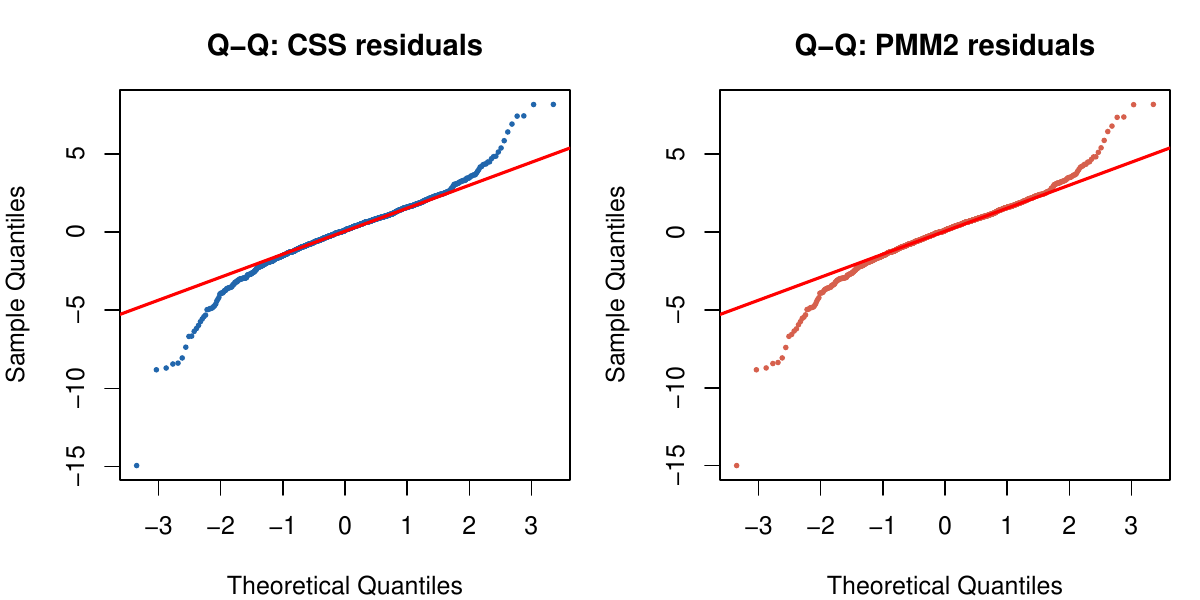}
\caption{Normal Q-Q plots of ARIMA(1,1,0) residuals: CSS-ML (left)
  and PMM2 (right).  Heavy tails confirm $|\gamma_3| > 0.5$, validating
  the PMM2 selection by \fct{pmm\_dispatch}.}
\label{fig:wti-qq}
\end{figure}

Both estimators are consistent; the AR(1) estimates differ by 0.013.
PMM2's BIC is 3.8 units lower, reflecting its smaller effective degrees of
freedom ($p = 1$ vs.\ CSS-ML's intercept-equivalent free parameter).
Figure~\ref{fig:wti-qq} shows that both residual distributions share a
similar, heavy-tailed non-Gaussian shape.

\subsection{Out-of-sample forecast evaluation} \label{sec:wti-oos}

An 80\%/20\% expanding-window 1-step-ahead evaluation (999 training,
250 test observations) yields Table~\ref{tab:wti-oos}.
The OOS gains are modest, as expected: with $g_2 \approx 0.93$ the
theoretical reduction in parameter asymptotic variance is~7\%, but
1-step-ahead forecast error is
dominated by the innovation variance $\sigma^2$, not by parameter uncertainty.
The value of PMM2 is therefore primarily in \emph{estimation precision}
(smaller confidence intervals, tighter AIC/BIC) rather than point forecast
accuracy.

\begin{table}[h!]
\centering
\caption{Out-of-sample 1-step-ahead forecast performance on WTI daily prices
  (250 test observations, 80/20 expanding-window split).
  RMSE and MAE in USD/barrel.} \label{tab:wti-oos}
\begin{tabular}{llrr}
\toprule
Model & Method & RMSE & MAE \\
\midrule
ARIMA(1,1,0) & CSS-ML & 1.371 & 1.021 \\
ARIMA(1,1,0) & PMM2   & 1.372 & 1.021 \\
\bottomrule
\end{tabular}
\end{table}

Full replication code is in \code{casestudy/wti\_jss\_condensed.R}.

%% -- 7. Auto MPG Case Study ---------------------------------------------------

\section[Auto MPG cross-sectional case study]{Auto MPG cross-sectional case study} \label{sec:autompg}

This section reproduces the two regression experiments of
\cite{zabolotnii2018polynomial} on the Auto MPG dataset
\citep{quinlan1993combining} from the UCI Machine Learning Repository
\citep{uciml}.  The dataset describes fuel consumption (\code{mpg}) and
seven other characteristics for $n = 392$ cars (after removing rows with
missing horsepower).  We use the version distributed in the
\pkg{ISLR2} package \citep{islr2}.  Both experiments fit a quadratic
polynomial
\[
\mathrm{MPG} \;=\; a_0 \,+\, a_1\, x \,+\, a_2\, x^2 \,+\, \varepsilon,
\]
but with two different predictors $x$ that produce qualitatively
different non-Gaussian residual structure --- the first asymmetric, the
second nearly symmetric and platykurtic --- exercising both branches of
the PMM dispatcher.

\subsection{MPG against vehicle weight (PMM2 branch)} \label{sec:autompg-weight}

We rescale weight from pounds to thousands of pounds (\code{weight\_klb})
to keep the polynomial design matrix well conditioned.  PMM2's
fixed-point iteration is otherwise sensitive to the
$\mathcal{O}(10^6)$ ratio between the linear and quadratic columns of
the raw-pound design matrix.

\begin{Code}
library("EstemPMM"); library("ISLR2")
dat <- na.omit(Auto)
dat$weight_klb  <- dat$weight / 1000
dat$weight_klb2 <- dat$weight_klb^2

fit_w_ols  <- lm(mpg ~ weight_klb + weight_klb2, data = dat)
res_w      <- residuals(fit_w_ols)
pmm_dispatch(res_w, verbose = TRUE)
## n = 392 | gamma3 = +0.809 | gamma4 = +1.770
##   g2(PMM2) = 0.826  |  g3(PMM3) = 0.861
##   >>> |gamma3| = 0.809 > 0.3 and g2 = 0.826 < 0.95:
##       moderate asymmetry, PMM2 worthwhile (17.4% reduction).

fit_w_pmm2 <- lm_pmm2(mpg ~ weight_klb + weight_klb2, data = dat)
\end{Code}

The OLS residuals are markedly asymmetric ($\hat\gamma_3 = 0.81$) and
moderately leptokurtic ($\hat\gamma_4 = 1.77$); the Jarque--Bera test
strongly rejects normality ($\chi^2_{(2)} = 93.9$, $p < 10^{-20}$).
The PMM2 efficiency coefficient $g_2 = 1 - \hat\gamma_3^2 /
(\hat\gamma_4 + 2) = 0.83$ predicts a 17\% reduction in asymptotic
variance.  Coefficient estimates and information criteria are reported
in Table~\ref{tab:autompg-weight}; the PMM2 estimates differ from OLS
mainly in the intercept (60.7 vs.\ 62.3) and the linear slope ($-17.96$
vs.\ $-18.50$ MPG per klb), reproducing the values in
\cite{zabolotnii2018polynomial} to within rounding.

\begin{table}[h!]
\centering
\caption{Auto MPG, regression on weight: OLS vs.\ PMM2.  Coefficients
  are for the rescaled predictor (klb); the published Zabolotnii et al.\
  values multiply $a_1$ by $10^{-3}$ and $a_2$ by $10^{-6}$.  AIC for
  PMM2 uses the package's S4 method; BIC is omitted because the current
  release does not implement a \fct{logLik} method on
  \class{PMM2fit}.} \label{tab:autompg-weight}
\begin{tabular}{lrrrrr}
\toprule
Method & $\hat a_0$ & $\hat a_1$ & $\hat a_2$ & AIC & BIC \\
\midrule
OLS  & $62.26$ & $-18.50$ & $1.697$ & $2238.1$ & $2254.0$ \\
PMM2 & $60.66$ & $-17.96$ & $1.696$ & $2240.7$ & --- \\
\bottomrule
\end{tabular}
\end{table}

\begin{figure}[h!]
\centering
\includegraphics[width=0.49\textwidth]{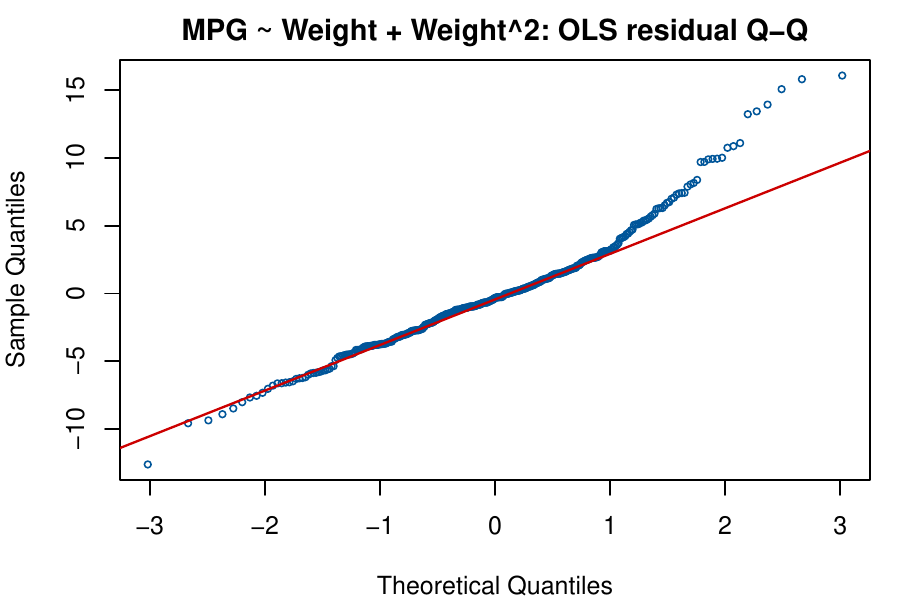}
\hfill
\includegraphics[width=0.49\textwidth]{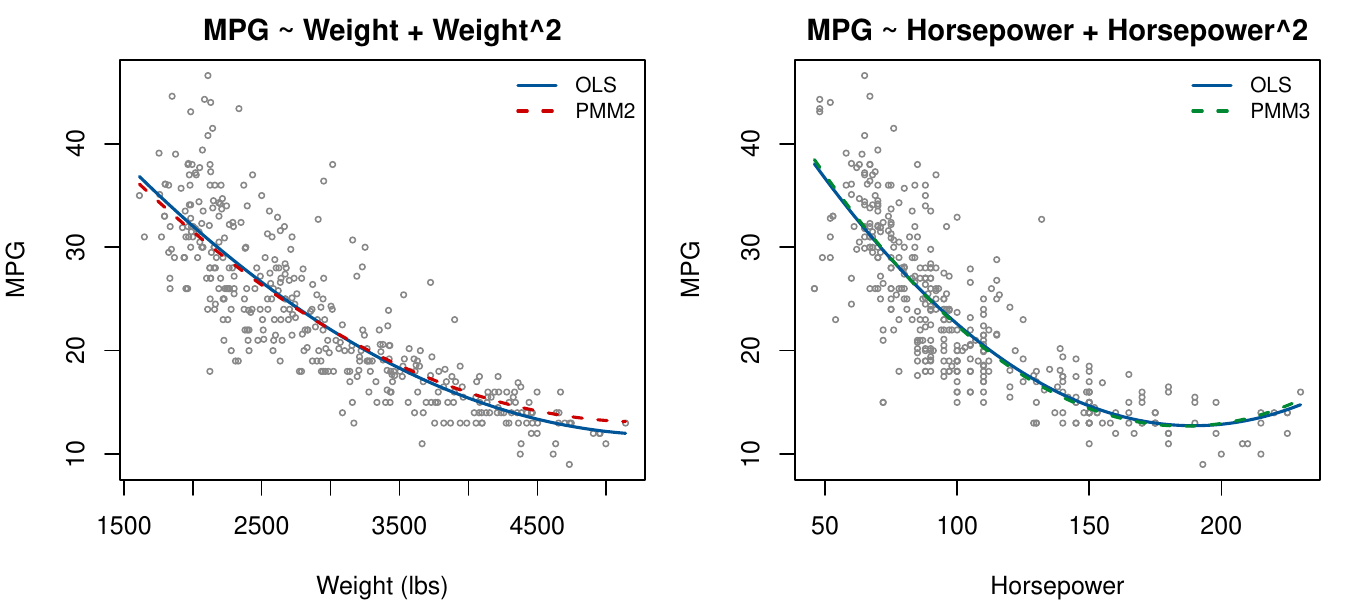}
\caption{Auto MPG --- weight regression.  Left: Q--Q plot of OLS
  residuals showing positive skew.  Right: scatter of MPG vs.\ weight
  with OLS (solid blue) and PMM2 (dashed red) quadratic fits; the two
  curves are visually similar but PMM2 has 17\% smaller asymptotic
  variance.  (Right panel also contains the horsepower model from
  Section~\ref{sec:autompg-horsepower}.)} \label{fig:autompg-weight}
\end{figure}

Figure~\ref{fig:autompg-weight} shows the OLS-residual Q--Q plot
(panel a) and the fitted curves overlaid on the data (panel b).  The
right tail of the Q--Q plot deviates clearly above the reference line,
confirming the positive skew that PMM2 exploits.

\subsection{MPG against horsepower (PMM3 branch)} \label{sec:autompg-horsepower}

Repeating the analysis with horsepower as predictor produces residuals
that are nearly symmetric ($\hat\gamma_3 = 0.22$, well below the
asymmetry threshold of $0.5$) but distinctly non-Gaussian in the
fourth and sixth cumulants ($\hat\gamma_4 = 1.30$, $\hat\gamma_6 =
-1.60$).  This is the regime that PMM2 cannot exploit but PMM3 can:

\begin{Code}
dat$hp100   <- dat$horsepower / 100
dat$hp100_2 <- dat$hp100^2

fit_h_ols  <- lm(mpg ~ hp100 + hp100_2, data = dat)
res_h      <- residuals(fit_h_ols)
pmm_dispatch(res_h, verbose = TRUE)
## n = 392 | gamma3 = +0.218 | gamma4 = +1.299
##   g2(PMM2) = 0.986  |  g3(PMM3) = 0.895
##   >>> gamma3 = 0.218, gamma4 = 1.299: near-Gaussian residuals.
##       No PMM advantage expected. Use OLS.

fit_h_pmm3 <- lm_pmm3(mpg ~ hp100 + hp100_2, data = dat)
\end{Code}

\textbf{Note on the dispatcher message.}  The verbose message from
\fct{pmm\_dispatch} is conservative: it suggests OLS because $g_2$
alone (the PMM2 coefficient) is close to $1$, even though the PMM3
coefficient $g_3 = 0.895$ predicts a 10.5\% asymptotic-variance
reduction over OLS.  The numerical comparison in
Table~\ref{tab:autompg-horsepower} shows that PMM3 produces a slightly
smaller AIC than OLS, consistent with the $g_3$ prediction; a more
discriminating dispatch rule that compares $g_3$ as well as $g_2$
against $1$ is on the package roadmap.

\begin{table}[h!]
\centering
\caption{Auto MPG, regression on horsepower: OLS vs.\ PMM3.
  Coefficients are for the rescaled predictor (hp/100).}
  \label{tab:autompg-horsepower}
\begin{tabular}{lrrrrr}
\toprule
Method & $\hat a_0$ & $\hat a_1$ & $\hat a_2$ & AIC & BIC \\
\midrule
OLS  & $56.90$ & $-46.62$ & $12.31$ & $2274.4$ & $2290.2$ \\
PMM3 & $58.18$ & $-48.90$ & $13.14$ & $2273.0$ & --- \\
\bottomrule
\end{tabular}
\end{table}

\begin{figure}[h!]
\centering
\includegraphics[width=0.6\textwidth]{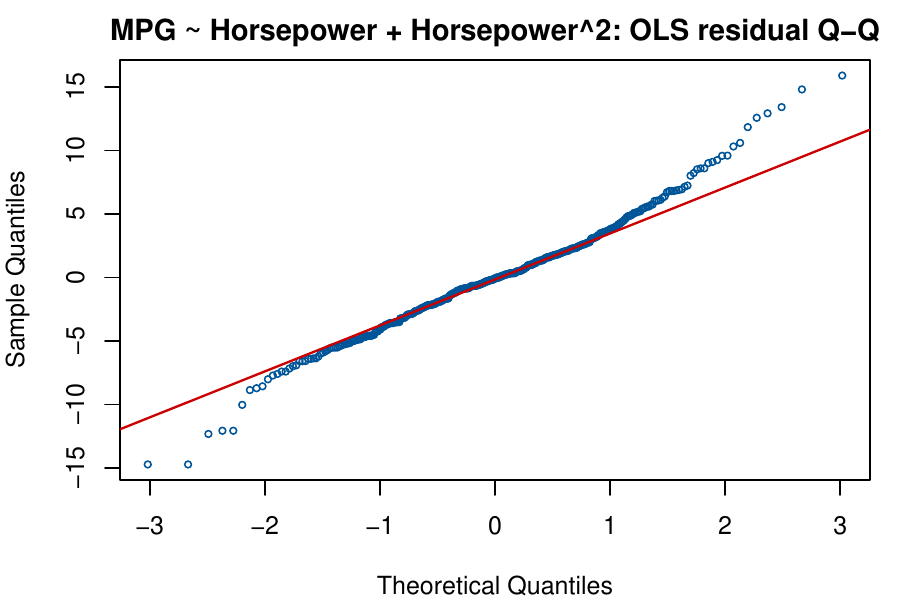}
\caption{Auto MPG --- horsepower regression: Q--Q plot of OLS
  residuals.  The two-sided tail flare reflects the platykurtic
  ($\hat\gamma_4 > 0$ but $\hat\gamma_6 < 0$) shape that PMM3 targets.}
\label{fig:autompg-horsepower}
\end{figure}

Figure~\ref{fig:autompg-horsepower} shows the Q--Q plot of the OLS
residuals: the points sit close to the reference line in the centre
but flare outward in both tails, the visual signature of a
platykurtic distribution.  The PMM3 fit produces coefficients that
agree with \cite{zabolotnii2018polynomial} (Experiment 2) to within
rounding.  Full replication code for both experiments is in
\code{casestudy/autompg\_jss.R}.

%% -- 8. Sunspot Case Study ----------------------------------------------------

\section[Wolfer sunspot AR(2) case study]{Wolfer sunspot AR(2) case study} \label{sec:sunspot}

The third case study illustrates PMM2 on a classical asymmetric
time-series benchmark.  The annual Wolfer sunspot series, distributed
in base \proglang{R} as \code{datasets::sunspot.year}, runs from 1700
to 1988 ($n = 289$).  \cite{box2015time} treat this series as a
canonical AR(2) benchmark; the same AR(2) order is selected by
\fct{ar} and \fct{auto.arima} when applied to it.

\begin{Code}
y       <- as.numeric(sunspot.year)
fit_css <- arima(y, order = c(2, 0, 0), method = "CSS-ML")
pmm_dispatch(residuals(fit_css), verbose = TRUE)
## n = 289 | gamma3 = +0.867 | gamma4 = +2.048
##   g2(PMM2) = 0.814  |  g3(PMM3) = 0.884
##   >>> |gamma3| = 0.867 > 0.3 and g2 = 0.814 < 0.95:
##       moderate asymmetry, PMM2 worthwhile (18.6% reduction).

fit_pmm2 <- ar_pmm2(y, order = 2)
\end{Code}

CSS residuals are strongly asymmetric ($\hat\gamma_3 = +0.87$) and
moderately leptokurtic ($\hat\gamma_4 = +2.05$); the Jarque--Bera
statistic is $86.7$ ($p < 10^{-18}$), and the theoretical efficiency
coefficient $g_2 = 1 - 0.867^2/(2 + 2.05) = 0.81$ predicts an 18.6\%
reduction in PMM2 asymptotic variance.  The two estimators yield
substantively different AR(2) coefficients (Table~\ref{tab:sunspot}):
PMM2's $\hat\phi_1 = 1.294$ and $\hat\phi_2 = -0.599$ both shrink
toward zero relative to CSS-ML's $1.389$ and $-0.691$, mildly
attenuating the cyclical persistence implied by the model.

\begin{table}[h!]
\centering
\caption{Sunspot annual AR(2) fits: CSS-ML vs.\ PMM2.  Block-bootstrap
  standard errors for PMM2 ($B = 500$, block length 11; one full
  solar cycle) are shown in parentheses.} \label{tab:sunspot}
\begin{tabular}{lrrr}
\toprule
Method & $\hat\phi_1$ & $\hat\phi_2$ & AIC \\
\midrule
CSS-ML & $+1.389$           & $-0.691$           & $2452.4$ \\
PMM2   & $+1.294\;(0.081)$  & $-0.599\;(0.064)$  & $2451.5$ \\
\bottomrule
\end{tabular}
\end{table}

\begin{figure}[h!]
\centering
\includegraphics[width=0.95\textwidth]{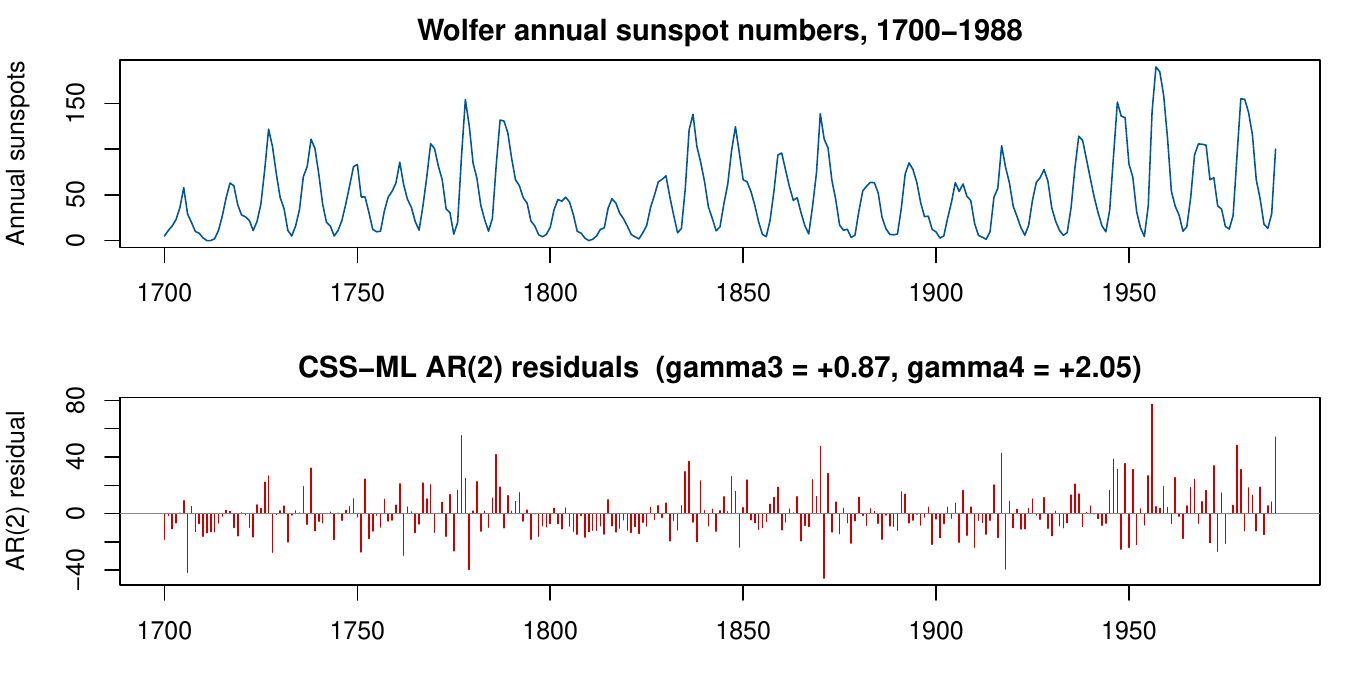}\\[2pt]
\includegraphics[width=0.55\textwidth]{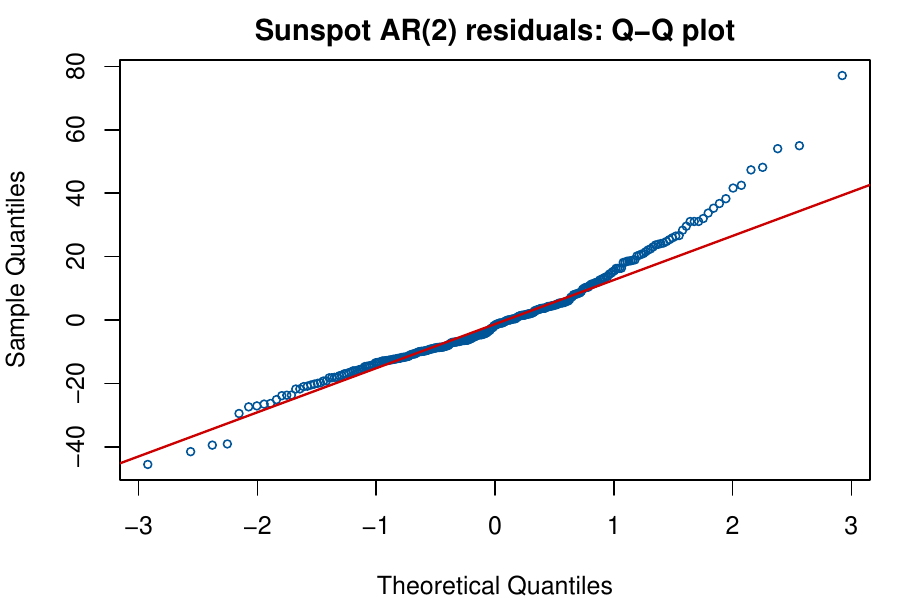}
\caption{Wolfer annual sunspot numbers (top), CSS-ML AR(2) residuals
  (middle), and Q--Q plot of those residuals (bottom).  The residual
  histogram is markedly right-skewed: the largest positive outliers
  reach $+50$ while the largest negative outliers stay near $-30$,
  yielding $\hat\gamma_3 \approx 0.87$ and $g_2 \approx 0.81$.}
\label{fig:sunspot}
\end{figure}

The PMM2 AIC is marginally lower ($2451.5$ vs.\ $2452.4$), and PMM2's
asymptotic variance reduction of 18.6\% is large enough to be
practically meaningful for confidence-interval calibration in
downstream cycle-period analyses.  Figure~\ref{fig:sunspot} shows the
series, the CSS residuals, and their Q--Q plot; the residuals'
right-skew --- positive outliers up to $+50$ versus negative outliers
bounded near $-30$ --- is the structural feature PMM2 exploits.  Full
replication code is in \code{casestudy/sunspot\_jss.R}.

%% -- 9. Summary ---------------------------------------------------------------

\section{Summary} \label{sec:summary}

\pkg{EstemPMM} provides a unified, production-quality \proglang{R}
implementation of the Polynomial Maximization Method for non-Gaussian linear
regression and ARIMA time-series models.  The package follows standard
\proglang{R} idioms (formula interface, S4 generics, \fct{AIC}/\fct{BIC}),
making it straightforward to substitute into existing workflows.

The primary recommendation for practice: apply \fct{pmm\_dispatch} to any
fitted model's residuals; if it selects PMM2 or PMM3, re-estimate with the
corresponding PMM function.  The additional computational cost is modest
(2--3$\times$ OLS for regression at $n = 200$, see Table~\ref{tab:cpu-lm}),
and the efficiency gains can be substantial when $|\gamma_3| \geq 0.5$.

\paragraph{Limitations.}
\begin{itemize}
  \item PMM2 and PMM3 require $|\gamma_3| \geq 0.5$ for a meaningful
        efficiency advantage; for mildly non-Gaussian data OLS/CSS suffices.
  \item Sample sizes $n < 200$ may yield noisy moment estimates that hurt
        rather than help PMM2 convergence.
  \item Seasonal ARIMA models with long periods ($s > 12$) may be slow to
        converge due to the large parameter space.
\end{itemize}

\paragraph{Future work.}
Extensions planned for future package versions include GARCH innovations
with PMM2 weighting, SARIMA with exogenous regressors (SARIMAX), and
parallel Monte Carlo via \pkg{parallel}.

%% -- Bibliography -------------------------------------------------------------

\bibliography{EstemPMM_JSS}

%% -- Appendix -----------------------------------------------------------------

\appendix

\section[R code for all examples]{R code for all examples} \label{app:code}

The file \code{code/all\_examples.R} contains all code from
Section~\ref{sec:illustrations} in a single self-contained script that can be
run as:

\begin{Code}
source(system.file("code", "all_examples.R", package = "EstemPMM"))
\end{Code}

\end{document}